\documentstyle[12pt,aasms4]{article}

\font\sevenbf=cmbx7
\def\note #1]{{\sevenbf #1 ------]}}
\def\eg{{\it e.g.~}}

\def\ie{{\it i.e.~}}

\def\arcsec{{\hbox{\rlap{\rlap{\tt\char"0D}\hbox{\thinspace\tt\char"0D}}
\kern-4.8pt\raise1pt\hbox{$\mit\mathchar"017F$}}}}
\def\arcmin{{\hbox{\rlap{\rlap{\tt\char"0D}}
\kern-5.5pt\raise1pt\hbox{$\mit\mathchar"017F$}}}}
\def\frac#1/#2{\leavevmode\kern.1em
\raise.5ex\hbox{\the\scriptfont0 #1}\kern-.1em
/\kern-0.15em\lower.25ex\hbox{\the\scriptfont0 #2}}
\def\jcd{Christensen-Dalsgaard}
\def\Fsurf{F_{\rm surf}}
\lefthead{Basu, D\"appen, \& Nayfonov}
\righthead{Helioseismic Analysis}

\begin{document}

\title{Helioseismic analysis of the hydrogen partition function
in the solar interior}

\author{Sarbani Basu\altaffilmark{1}, 
Werner D\"appen\altaffilmark{2,3}, 
and Alan Nayfonov\altaffilmark{2,4}} 

\altaffiltext{1}{Institute for Advanced Studies, 
School of Natural Sciences, Princeton, NJ 08540, U.S.A}
\altaffiltext{2}{Department of Physics and Astronomy, 
University of Southern California, Los Angeles, CA 90089-1342, U.S.A}
\altaffiltext{3}{Theoretical Astrophysics Center, Institute for
Physics and Astronomy, Aarhus Uni\-versity, 8000 Aarhus C, Denmark}
\altaffiltext{4}{IGPP, Lawrence Livermore National Laboratory,
Livermore, CA 94550, U.S.A.}

\begin{abstract} 
The difference in the adiabatic gradient $\gamma_1$ between inverted
solar data and solar
models is analyzed. To obtain deeper insight into the issues of plasma
physics, the so-called ``intrinsic'' difference in $\gamma_1$ is
extracted, that is, the difference due to the change in the equation of
state alone. Our method uses reference models based on two equations of 
state currently used in solar modeling, the 
Mihalas-Hummer-D\"appen (MHD) equation of state, 
and the OPAL equation of state (developed at
Livermore). Solar oscillation frequencies from the SOI/MDI instrument on
board the SOHO spacecraft during its first 144 days in operation are
used. Our results confirm the
existence of a subtle effect of the
excited states in hydrogen that was previously studied only
theoretically (\cite{nd98} 1998).
The effect stems from internal partition function of hydrogen, as 
used in the MHD equation of state. Although it is a pure-hydrogen
effect, it takes place in
somewhat deeper layers of the
Sun, where more than 90\% of hydrogen is ionized, and where
the second ionization zone of helium is located. Therefore,
the 
effect will have to be taken into account in reliable 
helioseismic determinations
of the astrophysically relevant
helium-abundance of the solar convection zone.
\end{abstract}


\section{Introduction}

Helioseismology has proved to be a an extremely powerful tool to study
the solar interior. Helioseismic inversion techniques have been used to
probe details of the solar structure (cf.~\cite{go96};~
\cite{bas96};~\cite{ko97}), abundances
(cf.~\cite{bant95};~\cite{ko96};~\cite{anchi98}), and rotation rate
(cf.~\cite{tho96};~\cite{sch98}). 
High-quality helioseismological data have allowed
us for some time 
to go beyond studying solar structure and probe the properties of
solar matter (see, \eg ,\cite{cd96}). Here, 
we try to study the equation of state of
solar matter using inversions of solar oscillation frequencies. 
Our inversions are differential, that is, they determine the solar
values in terms of deviations from given reference models.
For the present analysis two sets of reference models have been used.
The are based on the two equations of
state normally used to construct solar models (see section~2.2), the
so-called MHD equation of state 
(\cite{hm88} 1988;~\cite{mdh88};~\cite{dmhm88};~\cite{dam87}), 
and the OPAL equation of state (\cite{ro86};~\cite{ir95};~\cite{rsi96}
1996 and references therein), respectively.

The observed solar oscillation modes are standing acoustic waves; hence
the quantity most obviously probed is sound speed. Since the
oscillations are largely adiabatic (except very near the surface), the
frequencies are determined predominantly by the local adiabatic sound
speed $c$, which is a thermodynamic quantity. In addition, the
frequencies depend on the distribution of the density $\rho$ in the Sun. 
Equivalently, adiabatic oscillation frequencies can also be determined
from the distribution of density and the adiabatic gradient $\gamma_1$
in the Sun (see section~3.2 for more details). Since $\gamma_1$ is a
dimensionless quantity, it is especially suited to study qualitative
effects in the equation of state. In addition, for a high-precision
diagnosis of the equation of state it is very fortunate that the Sun has
a substantial convection zone, where equation of state effects show up
independent of our uncertainty in the opacity. The reason is that in the
bulk of the convection zone, the stratification is essentially adiabatic
and thus determined by thermodynamics (\cite{cd92}).

Helioseismic equation of state studies have shown the necessity to
include the leading Coulomb correction (expressed by the Debye-H\"uckel
approximation) in order to meet the helioseismological constraints. This
has been recognized in the early 80s and models with 
more sophisticated equations
of state were tried in helioseismic studies ({\it e.g.} \cite{be80};
\cite{l80}; \cite{ul82}; \cite{ur83}; \cite{sng83-84}; \cite{nsg84}).
With better data towards the end of the 80s, a clearer picture emerged
(\cite{cdl88} 1988; \cite{bant95} 1995; \cite{cd92}; \cite{cd96}). 
However, even those
equations of state that do include the leading Coulomb correction differ
from the solar data by more than the observational uncertainty. They
also differ discernibly among themselves, thus revealing the importance
of the treatment of the nonideal effects beyond the leading Coulomb
term. The aforementioned studies were based on the forward approach,
\ie , the comparison between observed and theoretically predicted solar
oscillation frequencies. Equation-of-state studies based on inversions
have also been attempted earlier 
(cf.~\cite{dps92};~\cite{el96};~\cite{ek98} 1998), but
they have been indirect. Those studies involved looking at either the
sound-speed differences or differences in the adiabatic index
$\gamma_1$, between the solar models and the Sun, both of which have
contributions from differences in structure and composition in addition
to contributions form the difference in the equation of state. 

In this paper we use the method proposed by ~\cite{bach97} (1997) to
study the difference in $\gamma_1$ between solar models that results
from the differences in the equation of state alone, but not from the
ensuing change in solar structure. To distinguish it from the total
$\gamma_1$ difference $\delta\gamma_1$, we call this contribution the
``intrinsic'' $\gamma_1$ difference $(\delta\gamma_1)_{\rm int}$.
An immediate target of the new
observational technique is 
the subtle effect of the
excited states in hydrogen that that has so far only been
studied theoretically (\cite{nd98} 1998).
We begin with a
presentation of the equation of state issues in order to
discuss the hydrogen
partition function effect. The solar oscillation
frequencies used for this work were obtained from the SOI/MDI instrument
on board the SOHO spacecraft during its first 144 days in operation
(\cite{rho97}). 

\section{Equation of State Issues}

\subsection{Ideal and Non-ideal Plasmas}

The simplest model is a mixture of nuclei and electrons, assumed fully
ionized and obeying the classical perfect gas law. However, an {\it
``ideal-gas''} equation of state can be more general. It may include
deviations from the perfect gas law, namely ionization, radiation and
degeneracy of electrons, as long as the underlying microphysics of these
additional effects is still ideal, that is, as long as it does not
contain interactions. The ``particles'', however, can be classical or
quantum, material or photonic. In such an ideal framework, bound systems
(molecules, atoms, ions) are allowed to have internal degrees of freedom
(excited states, spin). All such ideal effects can be calculated as
exactly as desired. Incidentally, in a recent study by~\cite{ek98}
(1998) it turned out that helioseismological accuracy demands inclusion
of the relativistic effect in the electron gas, which has been neglected
both in MHD and OPAL. Subsequently, both these equations of state are
now being upgraded to include a relativistic treatment of electrons. 
Interestingly, the simple but astrophysically useful~\cite{eff73}
equation of state (EFF) includes relativistic electrons.

One measure of nonideality in plasmas is the so-called coupling
parameter~$\Gamma$.
In a
plasma where particles have average distance $\langle r\rangle$ from
each other, we can define $\Gamma$ as the ratio of average potential
binding energy over mean kinetic energy $k_BT$ (in the simplest case of
hydrogen; generalizations to other elements are straightforward)

\begin{equation}
\Gamma = {{({e^2 \over \langle r\rangle})} \over {k_BT}}.
\label{e-bdn1}
\end{equation}

\noindent 
Plasmas with $\Gamma\gg 1$ are {\it strongly} coupled, those with
$\Gamma \ll 1$ are {\it weakly} coupled. A famous example of a strongly
coupled plasmas is the interior of white dwarfs, where the coupling can
become so strong as to force crystallization. Weakly coupled are the
interiors of stars with masses ranging from the slightly sub-solar ones
to the largest.

As one can suspect, $\Gamma$ is the dimensionless coupling parameter
according to which one can classify theories. Weakly-coupled plasmas
lend to systematic perturbative ideas ({\it e.g.} in powers of
$\Gamma$), strongly coupled plasma need more creative treatments.
Improvements in the equation of state beyond the model of a mixture of
ideal gases are difficult. This has both conceptual and technical
reasons. As a fundamental conceptual reason we mention the fact that in
a plasma environment already the idea of isolated atoms (and compound
ions) has to be abandoned. A technical reason is the difficulty
encountered when specific nonideal effect are modeled. The three
principal nonideal effects are related to: (i) the internal partition
functions of bound systems, (ii) pressure ionization, and (iii)
collective interactions of the charged particles. The internal partition
functions contain the difficult problem of excited states, where and how
they are to be cut off. They are an important element in determining the
ionization balances. Pressure ionization has to be provided by nonideal
interaction terms, because ideal gases would unphysically recombine in
the central regions of stars. 

\subsection{Chemical and Physical Picture}

There are two basic approaches to realize nonideal equations of state: 
the so-called {\it chemical} and {\it physical} pictures. In the
chemical picture one assumes that the notion of atoms and ions still
makes sense, and ionization is treated like a chemical reaction.
Modifications of atomic states by the surrounding plasma are expressed
in a heuristic and intuitive way. A typical procedure is to find an
expression for the probability that a given atomic state of an
individual atom still exists. Examples of equations of state realized in
the chemical picture are the MHD equation of state (see below) as well
as the aforementioned simple EFF equation of state.

The physical picture provides a systematic method to include nonideal
effects. The approach starts out from the grand canonical ensemble of a
system of the basic constituents (electrons and nuclei), interacting
through the Coulomb potential. Configurations corresponding to bound
combinations of electrons and nuclei, such as ions, atoms, and
molecules, arise in this ensemble naturally as terms in cluster
expansions. Any effects of the plasma environment on the internal states
are obtained directly from the statistical-mechanical analysis, rather
than by assertion as in the chemical picture. The only equation of state
realized in the 
physical picture that has been applied to stellar interiors
is the OPAL equation of state (see below). 

\subsubsection{Chemical Picture}

Most realistic equations of state that have appeared in the last 30
years belong to the chemical picture and are based on the free-energy
minimization method. This method uses approximate statistical mechanical
models (for example the nonrelativistic electron gas, Debye-H\"uckel
theory for ionic species, hard-core atoms to simulate pressure
ionization via configurational terms, quantum mechanical models of atoms
in perturbed fields, {{\it etc.}}). From these models a macroscopic free
energy is constructed as a function of temperature $T$, volume $V$, and
the particle numbers $N_1, \ldots, N_m$ of the $m$ components of the
plasma. This free energy is minimized subject to the stoichiometric
constraint. The solution of this minimum problem then gives both the
equilibrium concentrations and, if inserted in the free energy and its
derivatives, the equation of state and the thermodynamic quantities. 

Obviously, this procedure automatically guarantees thermodynamic
consistency. As an example, when the Coulomb pressure correction (to
the ideal-gas contribution) is taken into account in the free energy
(and not merely in the pressure), it affects both the pressure and the
equilibrium concentration, {{\it i.e.}}, the degrees of ionization. In
contrast, the mere inclusion of the pressure correction would be
inconsistent with other thermodynamic quantities. In the chemical
picture, perturbed atoms must be introduced on a more-or-less {\it
ad-hoc} basis to avoid the familiar divergence of internal partition
functions (see {\it e.g.} \cite{ekk76} 1976). In other words, the
approximation of unperturbed atoms precludes the application of standard
statistical mechanics, {\it i.e.} the attribution of a Boltzmann-factor
to each atomic state. The conventional remedy of the chemical picture
against this is a modification of the atomic states, {\it e.g.} by
cutting off the highly excited states in function of density and
temperature of the plasma. A currently popular equation of state
realized in the chemical picture is the MHD equation of state
(\cite{hm88} 1988;~\cite{mdh88};~\cite{dmhm88};~\cite{dam87}), 
based on an occupation probability
formalism. 
Specifically, the internal partition functions $Z_{s}^{\rm
int}$ of
species $s$ adopted by MHD are weighted sums

\begin{equation}
Z_{s}^{\rm int} = \sum_i w_{is }g_{is}
\exp \biggl( - { {E_{is} } \over k_B T} \biggr) \ .
\label{e-bdn2}
\end{equation}

\noindent
Here,the subscript $is$ labels states $i$ of species $s$. $E_{is}$ are their energies,
and the coefficients $w_{is}$ are the occupation probabilities that take
into account charged and neutral surrounding particles. In physical
terms, $w_{is}$ gives the fraction of all particles of species $s$ that
can exist in state $i$ with an electron bound to the atom or ion, and $1
- w_{is}$ gives the fraction of those that are so heavily perturbed by
nearby neighbors that their states are effectively destroyed.
Perturbations by neutral particles are based on an excluded-volume
treatment and perturbations by charges are calculated from a fit to a
quantum-mechanical Stark-ionization theory 
(for details see~\cite{hm88}
1988). 

\subsubsection{Physical Picture}

It is clear from the preceding subsection that the advantage of the
chemical picture lies in the possibility to model complicated plasmas,
and to obtain numerically smooth and consistent thermodynamic
quantities. Nevertheless, the heuristic method of the separation of the
atomic-physics problem from that of statistical mechanics is not
satisfactory, and attempts have been made to avoid the concept of a
perturbed atom in a plasma altogether. This has suggested an alternative
description, the physical picture. In such an approach one expects that
no assumptions about energy-level shifts or the convergence of internal
partition functions have to be made. On the contrary, properties of
energy levels and the partition functions should come out from the
formalism. 

There is an impressive body of literature on the physical picture.
Important sources of information with many references are the books
by~\cite{ekk76} (1976),~\cite{k86} (1986), and~\cite{eb91} (1991).
However, the majority of work on the physical picture was not dedicated
to the problem of obtaining a high-precision equation of state for
stellar interiors. Such an attempt was made for the first time by 
the OPAL group at
Livermore (\cite{ro86};~\cite{ir95};~\cite{rsi96}~1996 and references
therein). 

To explain the advantages of this approach for partially ionized
plasmas, it is instructive to discuss the activity expansion for gaseous
hydrogen. The interactions in this case are all short ranged and
pressure is determined from a self-consistent solution of the
equations~(\cite{ro81}) 

\begin{equation}
{p \over {k_BT}} = z + {z^2}{b_2} + {z^3}{b_3} + ...
\label{e-bdn3}
\end{equation}
\begin{equation}
\rho = {z \over {k_BT}} \biggl( {\partial p \over {\partial z} }
\biggr)\ ,
\label{e-bdn4}
\end{equation}

\noindent where $z={\lambda ^{-3}}{\exp (\mu /k_BT)}$ is the
activity, $\lambda \equiv h/\sqrt {2\pi m_e k_BT}$ is the thermal (de
Broglie) wavelength of electrons, $\mu$ is the chemical potential and T
is the temperature. The $b_n$ are cluster coefficients such that $b_2$
includes all two particle states, $b_3$ includes all three particle
states, {\it etc.}. 

In contrast to the chemical picture, which is plagued by divergent
partition functions, the physical picture has the power to avoid them
altogether. An important example of such a fictitious divergence is that
associated with the atomic partition function. This divergence is
fictitious in the sense that the bound-state part of $b_2$ is divergent
but the scattering state part, which is omitted in the Saha approach,
has a compensating divergence. Consequently the total $b_2$ does not
contain a divergence of this type (\cite{ekk76} 1976;~\cite{ro77}). A
major advantage of the physical picture is that it incorporates this
compensation at the outset. A further advantage is that no assumptions
about energy-level shifts have to be made (see the previous subsection);
it follows from the formalism that there are none. 

As a result, the Boltzmann sum appearing in the atomic (ionic) free
energy is replaced with the so-called Planck-Larkin partition function
(PLPF), given by ({\it e.g.}~\cite{ekk76} 1976;~\cite{k86}
1986;~\cite{ro86} and references therein) 

\begin{equation}
{\rm PLPF} =
\sum_{nl} (2l+1) \left[ \exp (-{E_{nl} \over k T} ) - \ 1 \ + {E_{nl} \over k
T} \right] \ .
\label{e-bdn5}
\end{equation}

\noindent
The PLPF is convergent without additional cut-off criteria as are
required in the chemical picture. We stress, however, that despite its
name the PLPF is not a partition function, but merely an auxiliary term
in a virial coefficient (see, for example, \cite{dam87}). 

\subsection{Coulomb Correction and the Debye-H\"uckel Approximation}

%%% do not format paragraph because of the \% signs!
%%% do not format paragraph because of the \% signs!
Debye-H\"uckel (DH) theory is based on the replacement of the long-range
Coulomb potential by a screened potential
(see {\it e.g.}~\cite{ekk76} 1976). This interaction leads
to a negative pressure correction to the ideal-gas value. Under solar
conditions, the relative correction culminates at about~8\% in the outer part
of the convection zone and it has another local maximum of about~1\% in the
core. Originally, the DH formalism included an additional ingredient of a
fixed size for positive ions, inside of which the electrostatic potential is
assumed to be constant. In astrophysical applications, such as MHD, this
so-called $\tau$-correction has been adopted in the form given by Harris {\it
et al.} (1960) (see also~\cite{g94};~\cite{ba96}). 
The systematic OPAL equation of state,
however, comes close to the unmodified DH results under solar condition.
This makes any
$\tau$-correction rather implausible, especially since
in comparisons with observational data, OPAL seems to fare better
than MHD (~\cite{cd96}), at least
below the helium ionization zones where perhaps other effects,
related to excited states, might
dominate (see section~4).

We would like to emphasize that close attention to the DH theory is
warranted, because it describes the main truly nonideal effect under
solar conditions. It was suggested in a number of early papers 
({\it e.g.}~\cite{be80}; 
\cite{ul82};~\cite{ur83};~\cite{sng83-84};~\cite{nsg84}) that
improvements in the equation of state can reduce discrepancies between
theory and observations. Later,~\cite{cdl88} (1988) showed that the MHD
equation of state reduced these discrepancies significantly for a large
range of oscillation modes. It turned out that the Coulomb term is the
dominant nonideal correction in the hydrogen and helium ionization
zones. This discovery led to an upgrade of the simple, but
astrophysically useful~\cite{eff73} equation of state (EFF) through the
inclusion of the Coulomb interaction term (CEFF)
(\cite{cd91};~\cite{cd92}). 

\subsection{Beyond the Debye-H\"uckel Correction}

Due to the relatively high temperature inside the Sun, the potential
energy of the Coulomb interaction is small compared to the kinetic
energy of particles, which allows us to believe that the DH theory makes
quantitative sense at least asymptotically. But to estimate the possible
error we need a physically based expression for next terms in the
corresponding expansion. The OPAL equation of state contains such higher
terms. An alternative is the path-integral based Feynman-Kac (FK)
formalism of~\cite{ap92} (1992, 1996),~\cite{acp94-95} (1994, 1995). Its
application to solar models is in progress~(\cite{pd98}). 

New insight beyond DH might come through the equivalent description of
the DH correction as a self energy of electrons and nuclei. A first
study has revealed that the thermodynamic consequence of screened
bound-state energies and a shifted continuum is hypersensitive to the
details of the screening and the method with which the thermodynamic
quantities are evaluated. No conclusion has been reached thus far but a
preliminary study has shown that static screening alone comes close to
the observational data, leaving little room for any thermodynamic
influence of dynamic screening (\cite{adn98}). Screening has also been
considered in connection with nuclear reaction rates, more specifically
those responsible for the solar neutrino flux ({\it e.g.}~\cite{bg97},
and references therein). 

Very recently,~\cite{nd98} (1998) examined the signature of the
internal partition function in the equation of state. That study has
revealed interesting features about excited states and their treatment
in the equation of state. The MHD equation of state with its specific,
density-dependent occupation probabilities (see section~2.2.1) is
causing a characteristic ``wiggle'' in the thermodynamic quantities,
most prominently in $\chi_{\rho} = (\partial \ln p / \partial \ln
\rho)_{T}$, but equally present in the other thermodynamic quantities. 

\placefigure{fig1}

Figures~1a and~1b show the presence of excited states in 
$\gamma_1$ for the case of a pure hydrogen plasma 
(panel {\it b} showing $\gamma_1$ relative to a simplified 
MHD equation of state containing only the ground-state 
contribution of hydrogen
atoms).
Temperatures and densities are taken from a solar model. Density is
implied but not shown in the figure (for more details and different
thermodynamic variables and chemical compositions, see \cite{nd98}
1998). Five cases were considered: (i) ${\rm MHD}$ [standard MHD
occupation probabilities of~\cite{hm88} (1988)], (ii) ${\rm MHD_{GS}}$
[standard MHD internal partition function of hydrogen but truncated to
the ground state (GS) term], (iii) ${\rm OPAL}$: OPAL tables [version of
November 1996 of~\cite{rsi96} (1996)], (iv) ${\rm MHD_{PL}}$ [MHD
internal partition function of hydrogen, but replaced by the
Planck-Larkin partition function, Eq.~(\ref{e-bdn5}), (v) ${\rm MHD_{PL,GS}}$
[${\rm MHD_{PL}}$ truncated to the ground state term]. The effect of the
inclusion of the excited states in the internal partition function is
manifest in the differences between MHD and ${\rm MHD_{GS}}$, and
between ${\rm MHD_{PL}}$ and ${\rm MHD_{PL,GS}}$, respectively. The
effect of the different occupation probability of the ground and excited
states shows up in the difference between MHD and ${\rm MHD_{PL}}$, and
between ${\rm MHD_{GS}}$ and ${\rm MHD_{PL,GS}}$. It was found that the
presence of excited states is crucial. Also, the wiggle, which is a
genuine neutral-hydrogen effect, is present despite the fact that most
of hydrogen is already ionized. The qualitative picture does not change
when helium is added (\cite {nd98} 1998). 

%\section{Inversion Technique}
\section{Inverting the Data}

\subsection{Introduction}

We have performed {\it differential} inversions with respect
to given reference models. We examine the influence of the equation
of state by choosing two sets of 
models, one with the MHD and the other with the
OPAL equation of state. Because of
its differential nature, our inversion
procedure is more reliable for reference models close
to the real solar structure. 
Since diffusion has now become part of the standard solar model
(\cite{cd96}), it is included in our reference models M1-M8
(see Table~1). For a study with artificial data, we have also
considered a reference model without diffusion (M9). We use
it to demonstrate that our inversion procedure applied
to the theoretical frequencies of 
model~M9 is
able to reconstruct the intrinsic $\gamma_1$-difference 
with respect to reference model (M5). We found that
this is indeed the case even for
the relatively bad model~M9, 
which is most likely farther
away from the true solar structure than 
each of the reference
models M1-M8 used for the real data.
In addition 
we have tested the robustness of the inferred results
despite uncertainties in solar model inputs by using a number
of different solar models.
Comparing the results
of inversions that are made with respect to different reference models
with the directly evaluated difference of the models themselves is a
powerful tool to assess the quality of the inversions.

\subsection{Method}

An inversion for solar structure 
(e.g.,~\cite{dps90};~\cite{da91};~\cite{anba94};~\cite{dzi94}) 
generally proceeds through a linearization of the equations
of stellar oscillations around a known reference model. The differences
between the structure of the Sun and the reference model are then
related to the differences in the frequencies of the Sun and the model
by kernels. Non-adiabatic effects and other errors in modeling the
surface layers give rise to frequency shifts 
(\cite{ck84};~\cite{balm92};~\cite{gue94};~\cite{ros95};~\cite{ros98})
which are not accounted for by the variational
principle. In the absence of any reliable formulation, these effects
have been taken into account in an {\it ad hoc} manner by including an
arbitrary function of frequency in the variational formulation
(\cite{dps90}). Thus the fractional change in
frequency of a
mode can be expressed in terms of fractional changes in the structure of
the model and a surface term. 

When the oscillation equation is linearized --- under the assumption of
hydrostatic equilibrium --- the fractional change in the frequency can
be related to the fractional changes in two of the functions that define
the structure of the models. Thus,

\begin{equation}
{\delta \omega_i \over \omega_i}
= \int K_{c^2,\rho}^i(r){ \delta c^2(r) \over c^2(r)}d r +
 \int K_{\rho,c^2}^i(r) {\delta \rho(r)\over \rho(r)} d r 
+{\Fsurf(\omega_i)\over E_i} \ .
\label{e-bdn6}
\end{equation}

\noindent 
(cf.~\cite{dps90}). Here $\delta \omega_i$ is
the
difference in the frequency $\omega_i$ of the $i$th mode between the
solar data and a reference model. The functions $c$ and $\rho$ are the
sound speed and density. The kernels $K_{c^2,\rho}^i$ and
$K_{\rho,c^2}^i$ are known functions of the reference model which relate
the changes in frequency to the changes in $c^2$ and $\rho$
respectively; and $E_i$ is the inertia of the mode, normalized by the
photospheric amplitude of the displacement. The term $\Fsurf$ results
from the near-surface errors. 

The conversion of the kernels for ($c^2$, $\rho$) to those of the 
set [$({\gamma_1})_{\rm int}$, $u$, $Y$], 
where, $u\equiv P/\rho$, $P$ being
the pressure and $Y$ the helium abundance is discussed
in~\cite{bach97} (1997). After the conversion, 
Eq.~(\ref{e-bdn6}) can be written
as 

\begin{equation}
{\delta
\omega_i \over \omega_i} = \int K_{(\gamma_1)_{\rm int}}^i { \left(\delta
\gamma_1 \over \gamma_1\right)}_{\rm int}d r + \int K_{u}^i{ \delta u
\over u}d r + \int K_{Y}^i {\delta Y} d r +{\Fsurf(\omega_i)\over
E_i}\ . 
\label{e-bdn7}
\end{equation}

We use the technique of Optimally Localized Averages (OLA) to carry out
the inversion. The purpose of the OLA technique 
(\cite{bg68};~\cite{ko95}) is to construct appropriate linear combinations,
$\sum_i c_i(r_0)\delta\omega_i/\omega_i$, of the data with coefficients
$c_i(r_0)$ so that the result represents a localized average of the 
quantity $(\delta \gamma_1 / \gamma_1)_{\rm int}$ at the radius $r_0$.
This is possible since the $(\gamma_1)_{\rm int}$ averaging kernel $\sum_i
c_i(r_0) K_{(\gamma_1)_{\rm int}}^i$ is a unimodular function localized
at $r=r_0$, and the sums $\sum_i c_i(r_0) K_{u}^i\delta u/u$, $\sum_i
c_i(r_0) K_{Y}^i\delta Y$ and $\sum_i c_i \Fsurf(\omega_i)/ E_i$ are
small. 

This objective is achieved by minimizing

\begin{eqnarray}
\int\left(\sum_i c_i K_{(\gamma_1)_{\rm int}}^i\right)^2(r-r_0)^2d r+
\beta_1\int \left(\sum_i c_i w(r)K_{u}^i\right)^2d r 
\nonumber \\
+\beta_2\int \left(\sum_i c_i w(r)K_{Y}^i\right)^2d r
+\mu\sum_{i,j}c_ic_j E_{ij} \; 
\nonumber \\
\label{e-bdn8}
\end{eqnarray}

\noindent
\noindent
with the constraint that the averaging kernel be uni\-modu\-lar, i.e.,

\begin{equation}
\sum_ic_i(r_0)\int K_{(\gamma_1)_{\rm int}}^i(r)d r =1 \; .
\label{e-bdn9}
\end{equation}

\noindent

Here, $E_{ij}$ is the covariance matrix of errors in the data. The
parameters $\beta_1$ and $\beta_2$ control the contributions of $\delta
u/u$ and $\delta Y$, respectively, and $\mu$ is a trade-off parameter
which controls the effect of data noise. The function $w(r)$ is a
suitably chosen, increasing function of radius, which ensures that the
contributions from the second and third terms from the surface layers
are suppressed properly. To reduce the influence of near-surface
uncertainties we apply the additional constraints that 

\begin{equation}
\sum_i 
c_i(r_0) E_i^{-1}\Phi_\lambda(\omega_i)=0 \; , \quad
\lambda=0,\ldots,\Lambda \; ,
\label{e-bdn10}
\end{equation}

\noindent 
where the $\Phi_\lambda$ are B-Splines with a suitably scaled argument
(cf.~\cite{da91}). 

\subsection{Reference Models}

We have constructed a set of calibrated solar models,
all of which have
either the MHD or the OPAL EOS. All models use OPAL opacities 
(\cite{ir96}) which are supplemented by the low temperature opacities
of~\cite{ku91} (1991). Since one of the most uncertain aspects of
solar
modeling is the formulation of the convective flux, we use two
different convective flux formalisms --- the standard mixing length
theory (MLT) and that of~\cite{cm91} (1991)
(henceforth called
CM). The two formalisms give fairly different stratifications in the
outer regions of the Sun where $(\gamma_1)_{\rm int}$ differences are
also likely to be the largest. All models have $Z/X=0.0245$
(cf.~\cite{gn93} 1993). 

\placetable{tbl-1}

A major source of uncertainty in the inversion results is the solar
radius. Till recently, the standard value of 695.99 Mm was used
unquestioningly. However, new f-mode data suggests that the radius 
actually smaller (cf.~\cite{sch97};~\cite{ant98}). A
difference in
the radius between the solar models used as a reference model for
inversions and the Sun can lead to a large error in the inversion
results (cf.~\cite{bas98a}). Thus we have constructed models
with the
standard radius as well as the f-mode radius of 695.78 Mm
(cf.~\cite{ant98}). 
Recent observations to determine the solar radius
(cf.~\cite{brch98}) suggest that the solar radius is even lower, 695.508 Mm.
Some of the relevant properties of the models are listed in 
Table~\ref{tbl-1}. All
models, except M9, assume the gravitational settling of helium and heavy
elements. 
The
composition profiles for  
models M1--M8 are the ones obtained from helioseismological inversions
by~\cite{anchi98} (1998). For model M9, the profile
from the no-diffusion model of~\cite{bp92} (1992) was used.

Note that the surface helium abundances and convection
zone depths of all
models except M9 are consistent with the helioseismically determined
helium abundance and depth of the convection zone of the solar envelope
(cf.~\cite{bant97} 1997; ~\cite{bas98a}).

\subsection{Test results}

To test the inversion procedure, we have inverted models M1 and M9 using
model M5 as the reference model. The purpose of inverting model M1 is to
show that we can invert for the intrinsic $\gamma_1$ differences between
two models (Fig.~2a). 
We see that
we are indeed successful in inverting for the intrinsic $\gamma_1$
difference between MHD and OPAL models. The structure and helium
abundance of the models M1 and M5 are very similar, hence the the
$\gamma_1$ difference between models with different equations of state
is dominated by differences in the EOS rather than differences in
structure. 

\placefigure{fig2}

In a parallel test, 
model~M9 is inverted with respect to reference model~M5 in order 
to show that we are indeed inverting just for
the 
equation of state part of 
the $\gamma_1$-difference $(\delta\gamma_1)_{\rm int}$.
We deliberately choose for this test the 
relatively bad solar model~M9 without diffusion, to
show insensitivity to the quality of the reference model. Importantly
for this test, model~M9 has the same
equation of state as the reference model~M5.
Using theoretically computed frequencies of model~M9, our inversion
procedure gives the intrinsic
difference of $\gamma_1$ between models~M5 and~M9.
Since the equation of state is the same in the two models,
we should obtain essentially zero for the intrinsic difference. This is
indeed the case (see Fig~2b).

The small deviations
from zero can be explained mainly as contributions 
in the solution because
of the large difference in the helium abundance [i.e. the third term in
Eq.~(\ref{e-bdn7})], though there can also be a contribution from the
neglected
differences in the heavy-element abundances. Increasing the parameter
$\beta_2$ helps in getting better inversion results, 
but it degrades the quality of the inversion. However, this is
not a concern for inversion of the solar data with models M1--M8, since
the difference in the 
helium abundance between these models and the Sun is
an order of magnitude less than the helium-abundance difference between
models M9 and M5. 
For comparison, Fig.~2b also shows
total $\gamma_1$ difference. 
Note that the total $\gamma_1$ inversion captures the difference in
total $\gamma_1$ to within the limits of the resolution.

\placefigure{fig3}

The mode set and errors used for the test were the same as that of the
SOI/MDI set mentioned above (\cite{rho97}). 
We are quite successful in constructing
localized kernels till fairly close to the solar surface. A sample of
the averaging kernels is shown in Fig.~3. The use of the observational
mode-set and errors ensures that the inversion properties ({\it e.g.}
resolution of the averaging kernels, propagated errors {\it etc.}) 
will remain
the same when solar data are inverted. 

\section{Results and Conclusion}

The result obtained by inverting solar data using models M1--M8 are
shown in Fig.~4.
We show inversions with respect to a
deliberately large number of reference models, having different
parameters. By doing so we demonstrate that in the chosen technique
only the equation of state matters.

Note that the results of all the MHD models are very similar despite
differences in radius and stratification. However these are very
different from the OPAL models which form a distinct group. The
differences between the two equations of state and the Sun are fairly
large in the outer layers of the Sun

\placefigure{fig4}

%%%do not justify (% signs!)  
%%%do not justify (% signs!)  
In contrast to earlier
results (~\cite{bach97} 1997), which could only resolve the somewhat
deeper layers of 
the Sun, the present focus on the 20\% uppermost layers confirms
the aforementioned effect of excited
states as included in the MHD equation of state (section~2.4). 
The results also confirm that the OPAL
equation of state is better for $r < 0.97 R_{\odot}$
(\cite{cd96}),
but the situation is
reversed in the top 2-3 per cent of the radius.

Since the difference in $\gamma_1$ between MHD
and OPAL is the wiggle of Fig.~1, the observed preference of the MHD
model in the upper region could indicate the correctness of an MHD-like
treatment of the exited states. Again, we emphasize that
Fig.~4 confirms earlier results that
below the wiggle region, OPAL fares better than MHD~(\cite{cd96}).
There, the internal partition functions of OPAL {\it \`a la}
Planck-Larkin, and the absence of a $\tau$-correction (section~2.3)
seem to be the better choice. 

The reversal of fortune in favor of MHD in the upper part of the Sun
(above~0.97 solar radii) could be due to the different implementations
of many-body interactions in the two formalisms. Since density
decreases in the upper part, OPAL by its nature of a systematic
expansion, inevitably becomes more accurate; but MHD might, by
its heuristic approach (and by luck!), have incorporated even finer,
higher-order effects. 

The helioseismic helium-abundance determination
procedure is
quite sensitive to the details of internal partition functions. Since
the wiggle is located in the second ionization zone helium, although it
is a pure hydrogen effect (see section~2.4), it has a very important
bearing on the astrophysically relevant helioseismic helium-abundance
$Y$
determination in the solar convection zone.
In principle, the asymptotic analysis by~\cite{bant95} (1995), which 
leads to $Y=0.249$ for
OPAL and $Y=0.246$ for MHD, could be taken as indication
that the equation of state does not matter in the result
for~$Y$. However,
other studies came to a different conclusion. In independent 
inversions,~\cite{kos97} (1997)
obtained $=0.23$ for MHD and
$Y=0.25$ for OPAL,~\cite{ric98} (1998) $Y=0.242$
for MHD and $Y=0.248$ for OPAL,
and~\cite{bas98b} (1998b) $Y=0.248$ for OPAL and 
$Y= 0.251$ for MHD. In an entropy calibration by~\cite{bayu97}
(1997) the result was $Y=0.23$ for OPAL
and $Y=0.25$ for MHD. In the latter two, the
sign of the change is reversed compared to the three former.

The confirmation of the excited-states effect in solar data
makes helium-abundance determinations
based on the MHD equation of state more plausible
than the ones based on OPAL, but the various current techniques
do not yet allow a consensus. We note in passing that
the helium-abundance values
given in Table~1 
should not be confused with the aforementioned inversion
results. Table~1 lists the result of
calibrations in which the helium parameter
compensates for other effects as well, and the relative
stability of the helium parameter in~M1--M8 is no indication
for its absolute correctness. Inversions can in principle
isolate the
helium abundance from other effects, but an absolutely
accurate equation of state is required. As we mentioned,
any quantitative conclusions about the value of $Y$ 
would be premature. However, since the 
MHD equation of state appears to describe reality in the
helium-ionization zone better, 
it might be the preferable equation of state
in helium-abundance determinations. 

Let us add a word of caution, though. It could appear tempting to
produce a ``combined'' solar equation of state, with MHD for the top
part and OPAL for the lower part. However, such a hybrid solution is
fraught with danger. For instance, it is known that patching together
equations of state can introduce spurious effects (\cite{da93}). It
seems that the right way is to improve MHD and OPAL in parallel and
independently, guided by the progress of helioseismology. 

\acknowledgments

We thank J\o rgen Christensen--Dalsgaard and 
Forrest Rogers for stimulating discussions and valuable advice. A.N.
and W.D. are
supported by the grant AST-9618549 of the National Science
Foundation. W.D. acknowledges additional support from
the SOHO Guest Investigator grant NAG5-6216 of NASA,
a grant extended to the
University of Southern California by the Theory Group of Lawrence Livermore
National Laboratory, and
from the Danish National Research
Foundation through its establishment of the Theoretical Astrophysics Center.
S.B. is
supported by funds from the Institute for Advanced Study. 
SOHO
is a project of international cooperation between ESA and NASA.

\clearpage

{\bf FIGURE CAPTIONS}

\figcaption{
{\it Panel (a):} absolute values of $\gamma_1$ for solar temperatures
and densities of a hydrogen-only plasma. Linestyles: MHD -- asterisks,
${\rm MHD_{GS}}$ -- dashed lines, ${\rm MHD_{PL}}$ -- dotted-dashed
lines, ${\rm MHD_{PL,GS}}$ -- dotted lines, and OPAL -- solid lines. 
{\it Panel (b):} relative differences with respect to ${\rm MHD_{GS}}$,
in the sense $(\gamma_1 - \gamma_1[{\rm MHD_{GS}}])/\gamma_1[{\rm
MHD_{GS}}])$, using the same line styles as in (a). The horizontal
solid zero line, representing ${\rm MHD_{GS}}$, is also shown. See text
for the definitions of the different MHD versions. 
\label{fig1}}

\figcaption{
{\it Panel (a):} Result of inverting model M1 using model M5
as the reference model. The continuous line is the exact difference
in the intrinsic $\gamma_1$ between the two models. The points are the
results obtained by inverting the frequency differences between the
two models. The vertical error bars represent the errors in the
solution because of observational errors in the frequencies. The
horizontal error bars are a measure of the resolution of the 
inversion and is the distance between the quartile points of the
averaging kernels.
{\it Panel (b):} Result of inverting model M9 with model
M5 to obtain the intrinsic and total $\gamma_1$ difference.
The continuous line is the exact difference in total $\gamma_1$
between the two models. The models were constructed with the same
EOS, hence the intrinsic $\gamma_1$ difference is zero. 
\label{fig2}}

\figcaption{
Averaging kernels obtained at selected radii.
\label{fig3}}

\figcaption{
Relative difference between $\gamma_1$ obtained from an inversion of
helioseismological data and $\gamma_1$ of the solar models M1--M8 listed
in Table~I, in the sense ``Sun -- model''. Only the ``intrinsic''
difference in $\gamma_1$ is shown, that is, the part of the difference
due to the equation of state (see text). The filled points are results
obtained with MHD models, the empty ones are results with OPAL models.
Lines have been drawn through results of models M1 and M5 to guide the
eye. Error bars have been drawn only on two sets of results for the sake
of clarity.
\label{fig4}}

\clearpage

\plottwo{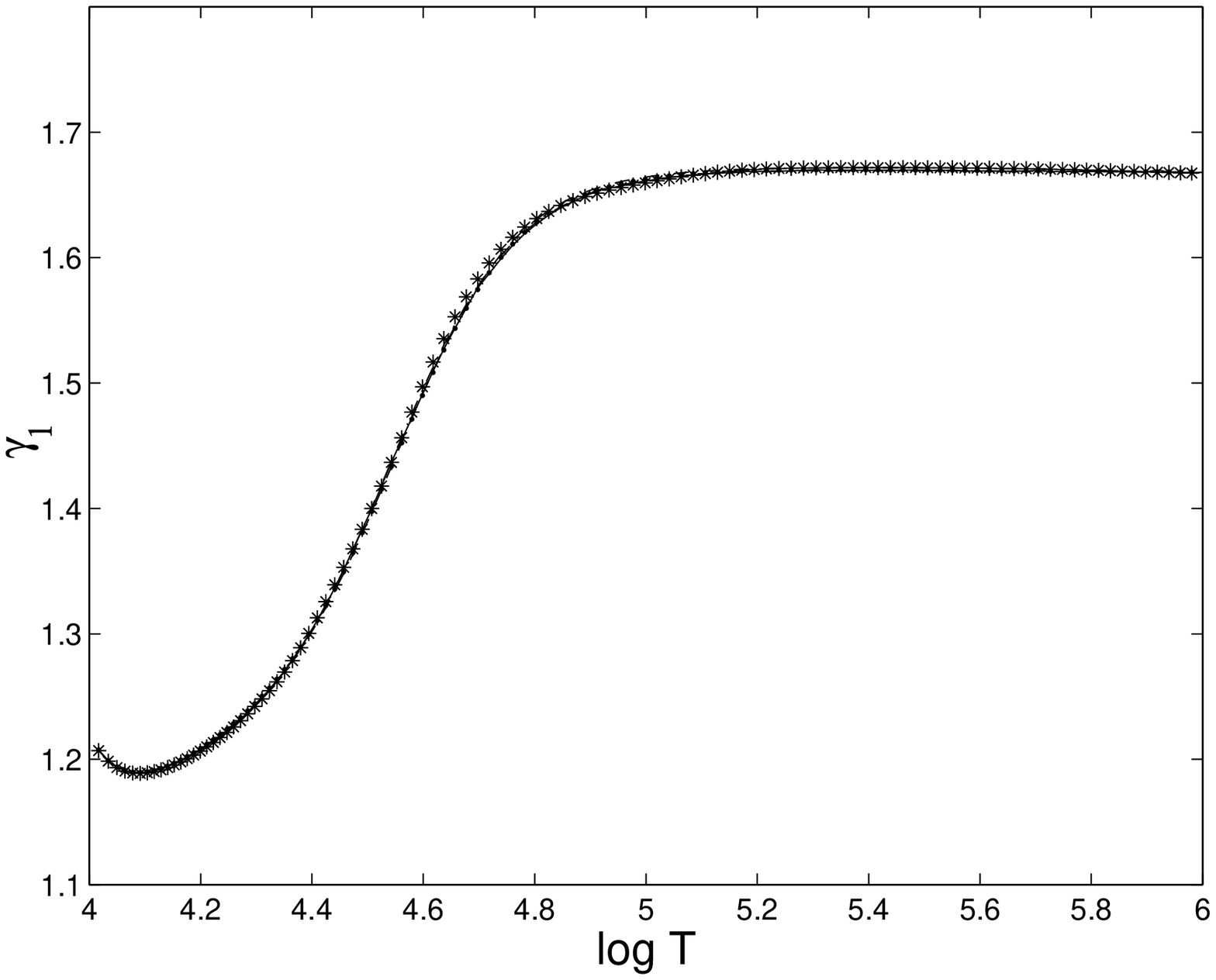}{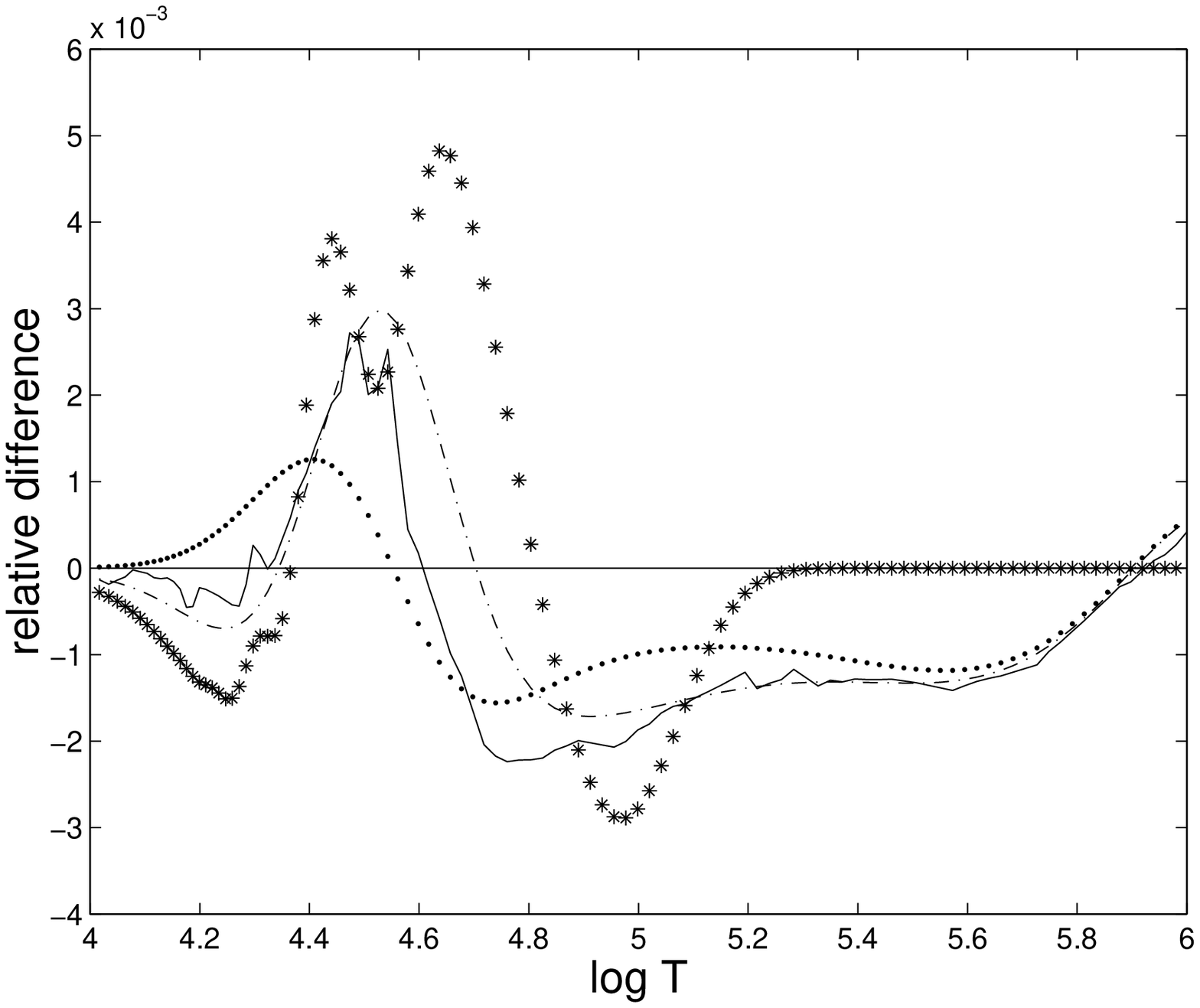}

\clearpage

\plottwo{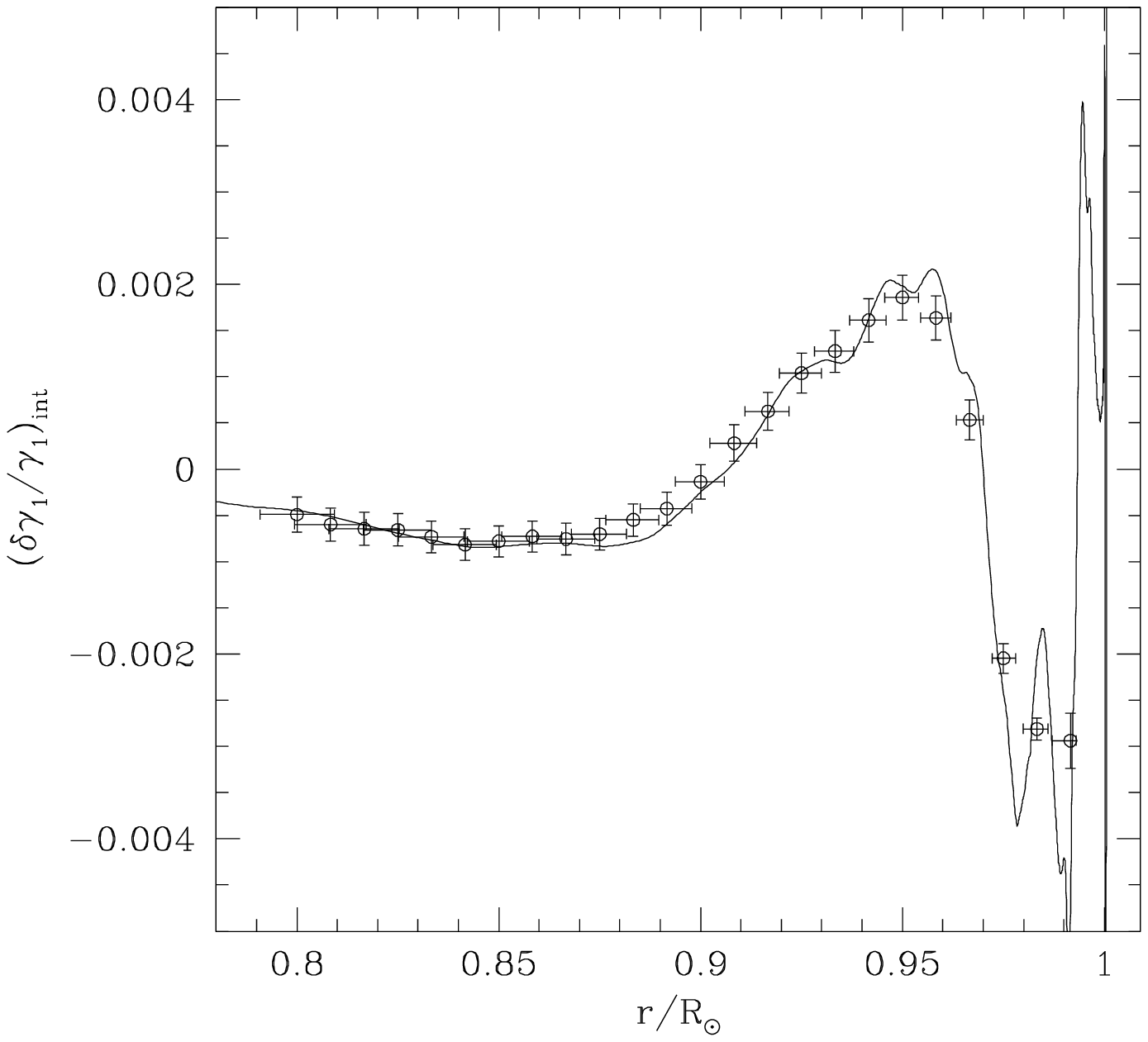}{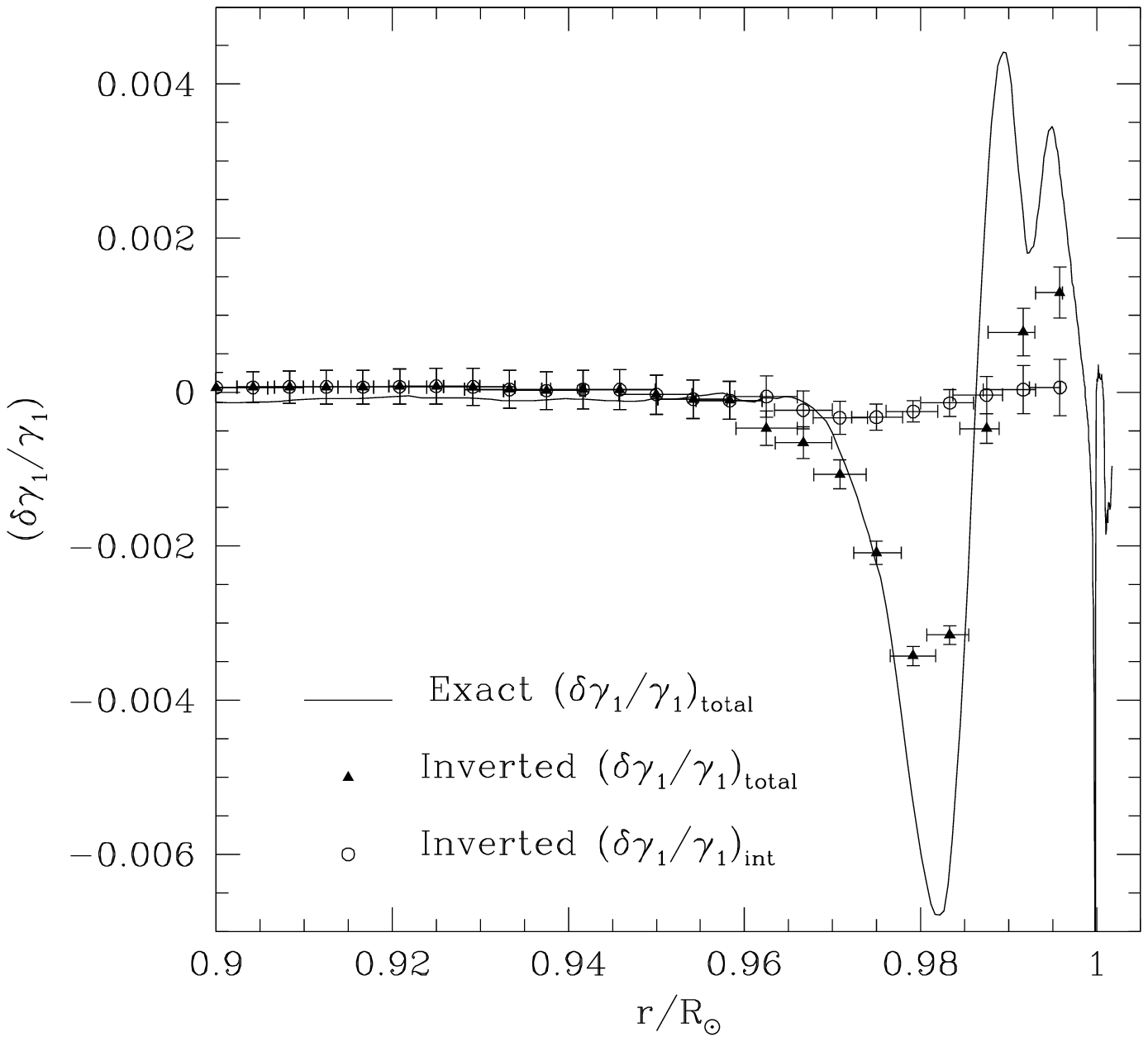}

\clearpage

\plotone{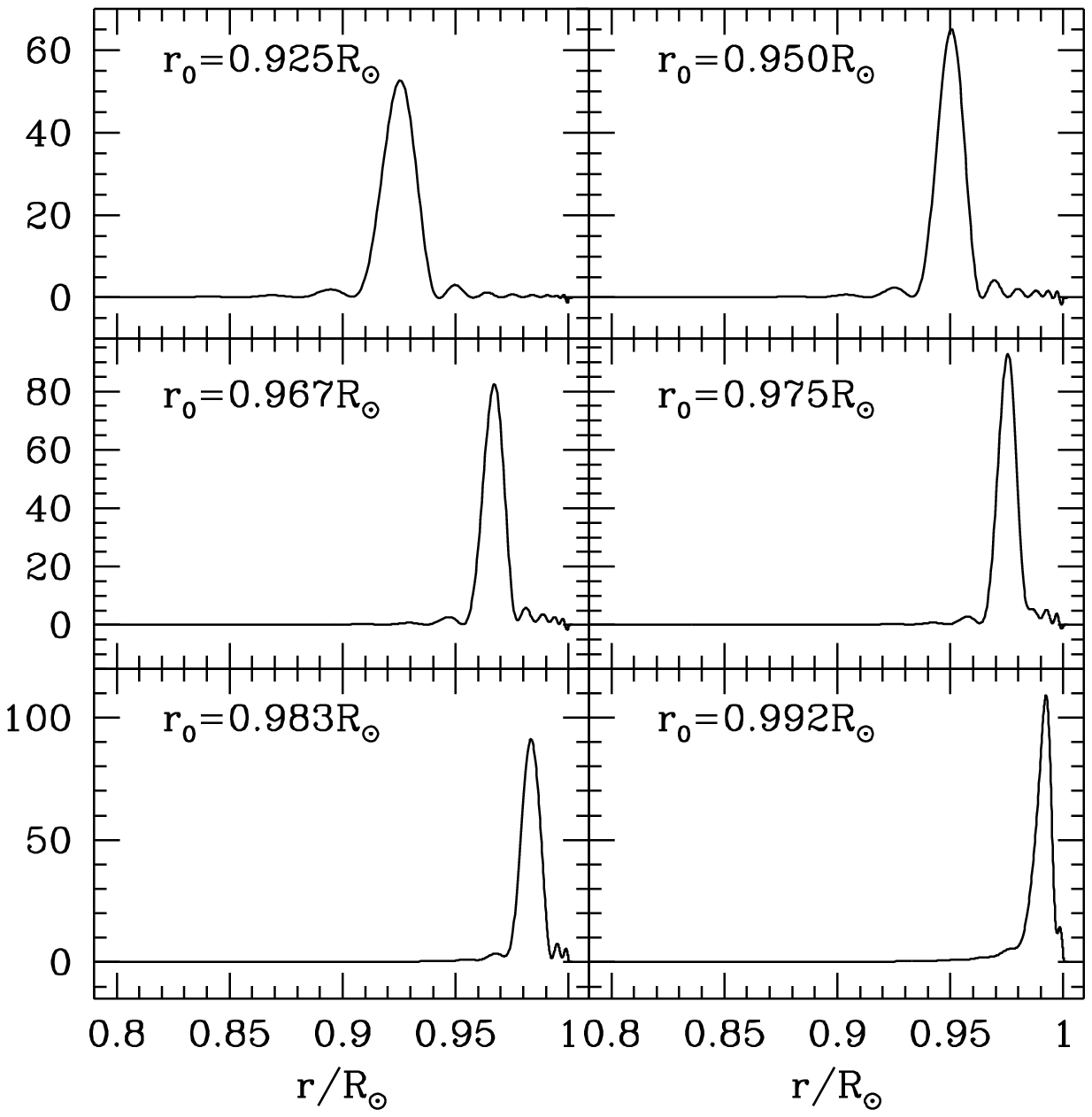}

\clearpage

\plotone{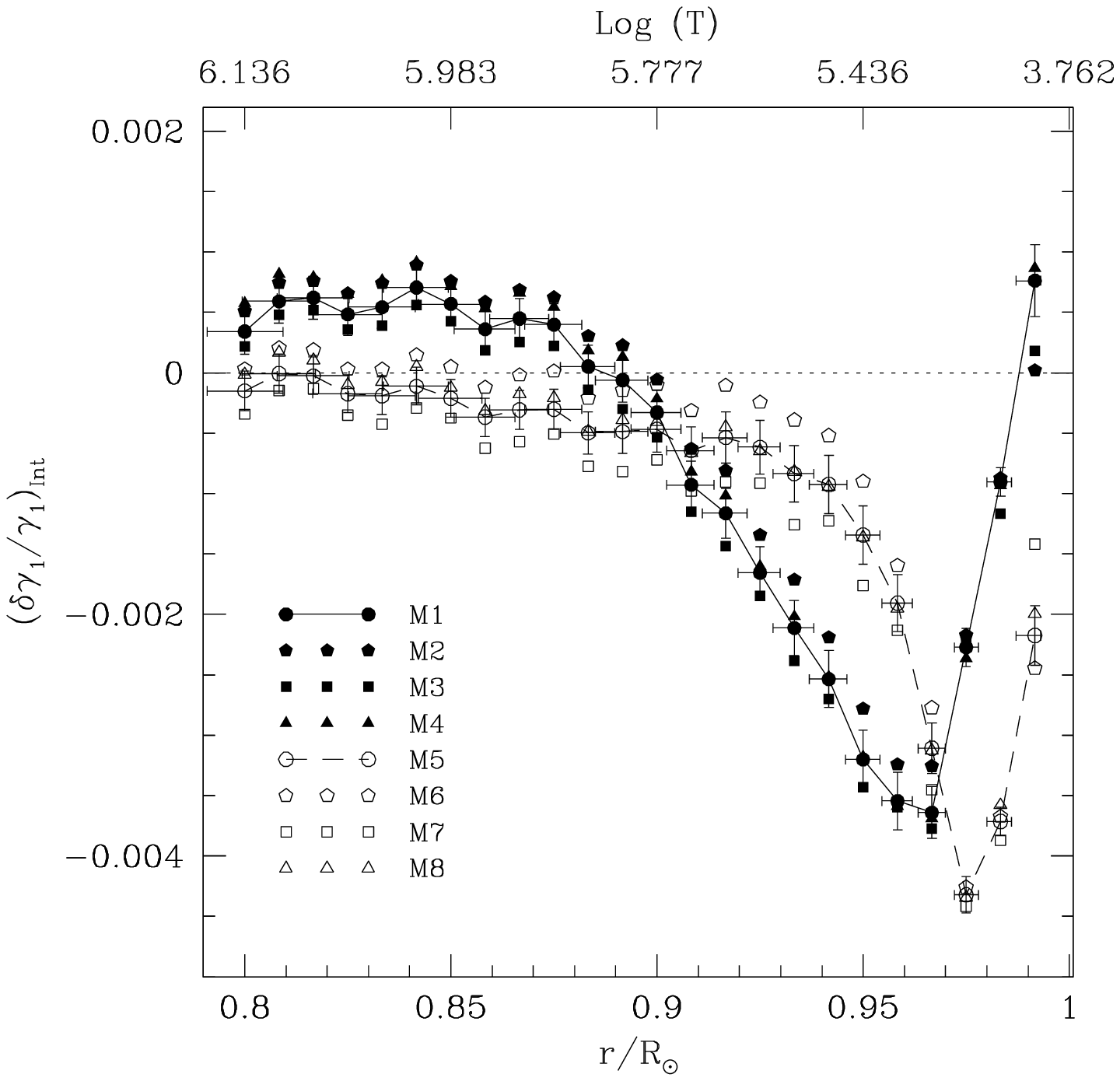}

\clearpage

\begin{table}

\caption{Properties of the solar models used in Figs.~2a,~2b, and~4 (see
text). 
\label{tbl-1}}
\vskip 1cm
\begin{tabular}{llllll}
\hline

Model & \hbox{EOS} & \hbox{Radius} & \hbox{Convective}& 
      $Y_s$& $r_{\rm cz}/R_\odot$ \\
      & & \hbox{Mm} & \hbox{Flux}& & \\

\hline

M1 & MHD & 695.78 & CM & 0.2472 & 0.7145 \\
M2 & MHD & 695.99 & CM & 0.2472 & 0.7146 \\
M3 & MHD & 695.51 & CM & 0.2472 & 0.7145 \\
M4 & MHD & 695.78 & MLT & 0.2472 & 0.7146 \\
M5 & OPAL& 695.78 & CM & 0.2465 & 0.7134 \\
M6 & OPAL& 695.99 & CM & 0.2465 & 0.7135 \\
M7 & OPAL& 695.51 & CM & 0.2466 & 0.7133 \\
M8 & OPAL& 695.78 & MLT & 0.2465 & 0.7135 \\
M9 & OPAL& 695.78 & CM & 0.2646 & 0.7268 \\

\hline
\end{tabular}
\end{table}


\clearpage
 
\begin{thebibliography}{}

\bibitem[Alastuey \& Perez]{ap92}
Alastuey, A., \& Perez, A. 1992,
%{``Virial Expansion of the Equation of State of a Quantum Plasma''},
Europhys.Lett., 20, 19-24

\bibitem[Alastuey \& Perez 1996]{ap96} 
Alastuey, A., \& Perez, A. 1996,
%{``Virial Expansion for Quantum Plasmas, Exchange Contributions''},
Phys. Rev. E, 53, 5714

\bibitem[Alastuey et al.]{acp94-95} 
Alastuey, A., Cornu, F., \& Perez, A. 1994,
%{``Virial Expansion for Quantum Plasmas, Diagrammatic Resummations''}
Phys. Rev. E, 49, 1077

%---------------- this ref is being cited together with the one above
\bibitem[]{acp95} 
Alastuey, A., Cornu, F., \& Perez, A. 1995,
%{``Virial Expansion for Quantum Plasmas, Maxwell-Boltzmann Statistics''},
Phys. Rev. E, 51, 1725

\bibitem[Antia 1998]{ant98}Antia H. M. 1998, A\&A, 330, 336

\bibitem[Antia \& Basu 1994]{anba94}
Antia H. M., \& Basu S. 1994, A\&AS, 107, 421

\bibitem[Antia \& Chitre]{anchi98}Antia H. M., \&
Chitre S. M. 1998, A\&A, in press (astro-ph/9710159)

\bibitem[Arndt et al. 1998]{adn98}
Arndt, A., D\"appen, W., \& Nayfonov, A. 1998,
\apj\ , 498, 349

\bibitem[Backus \& Gilbert 1968]{bg68}
Backus, G., \& Gilbert, F. 1968, Geophys. J. R. Astr. Soc., 16, 169

\bibitem[Bahcall \& Pinsonneault]{bp92}
Bahcall, J. N., \& Pinsonneault, M. H. 1992, Rev. Mod. Phys., 64, 885

\bibitem[Balmforth 1992]{balm92}Balmforth N. J. 1992, MNRAS, 255, 632

\bibitem[Basu 1998a]{bas98a}Basu, S. 1998a, MNRAS, 296, 1137

\bibitem[Basu]{bas98b}Basu, S. 1998b, MNRAS, 298, 719

\bibitem[Basu \& Antia]{bant95}Basu S., Antia H. M. 1995, MNRAS,
276, 1402

\bibitem[ Basu \& Antia]{bant97}Basu S., \& Antia H. M. 1997, MNRAS,
287, 189

\bibitem[Basu \& \jcd ]{bach97}Basu S., \& \jcd , J. 1997, 
A\&A, 322, L5

\bibitem[Basu et al.~1998]{bdn98}
Basu, S., D\"appen, W., \& Nayfonov, A. 1998, 
in Proc. SOHO6-GONG98 Workshop, ed.
S. Korzennik (ESA-SP), in press

\bibitem[ Basu et al.~1996]{bas96}Basu S., 
Christensen-Dalsgaard J., Schou J., Thompson M.J., \&
Tomczyk S. 1996, ApJ, 460, 1064

\bibitem[Baturin \& Ayukov]{bayu97}
Baturin, V.A., \& Ayukov, S.V. 1997,
in SCORe'96: Solar Convection and Oscillations and their Relationship,
ed. F.P. Pijpers, J. Christensen--Dalsgaard \& C. Rosenthal 
(Dordrecht: Kluwer), 55

\bibitem[Baturin et al. 1996]{ba96}
Baturin, V.A., D\"appen, W., Wang, X., \& Yang, F. 1996, 
in Proc. 32nd Li\`ege International Astrophysical Colloquium ``Stellar
Evolution: What should be done'', ed.
M. Gabriel \& A. Noels (Li\`ege: Institut d'Astrophysique), 33

\bibitem[Berthomieu et al. 1980]{be80}
Berthomieu, G., Cooper, A.J., Gough, D.O., Osaki, Y.,
Provost, J., \& Rocca, A. 1980,
%[Sensitivity of five minute eigenfrequencies to the structure of the Sun].
in Lecture Notes in Physics,
Vol. 125: Nonradial and Nonlinear Stellar Pulsation,
ed. Hill, H.A. \& Dziembowski, W. (Berlin: Springer), 307

\bibitem[Brown \& \jcd ~1998]{brch98}Brown, T., \& \jcd, J. 1998, 
ApJ, 500, L195

\bibitem[Br\"uggen \& Gough 1997]
{bg97}
Br\"uggen, M., \& Gough, D.O. 1997, ApJ, 488, 867

\bibitem[Canuto \& Mazitelli]{cm91}
Canuto, V. M., \& Mazzitelli, I. 1991, ApJ, 370, 295

\bibitem[Christensen-Dalsgaard 1991]{cd91}
Christensen-Dalsgaard, J. 1991,
%[Solar oscillations and the physics of the solar interior].
In
Lecture Notes in Physics, Vol. 388:
{\rm Challenges to Theories of the Structure of Moderate-mass Stars},
ed. D.O. Gough \& J. Toomre (Heidelberg: Springer), 11

\bibitem[Christensen-Dalsgaard \& D\"appen 1992]{cd92} 
Christensen-Dalsgaard, J., \& D\"appen, W. 1992,
Astron. Astrophys. Review, 4, 267

\bibitem[Christensen-Dalsgaard et al.]{cdl88} Christensen-Dalsgaard, J.,
D\"appen, W., \& Lebreton, L. 1988, 
Nature, 336, 634

\bibitem[Christensen-Dalsgaard et al. 1996]{cd96} 
Christensen-Dalsgaard, J., D\"appen, W., \& the GONG Team 1996,
Science, 272, 1286

\bibitem[Cox \& Kidman 1984]{ck84}
Cox A. N., \& Kidman R. B. 1986, in Theoretical Problems in
Stellar Stability and Oscillations (Li\`ege: Institut d'Astrophysique),
259

\bibitem[D\"appen et al. 1987]{dam87}
D\"appen, W., Anderson, L.S., \& Mihalas, D. 1987,
ApJ, 319, 195

\bibitem[D\"appen et al. 1988]{dmhm88} D\"appen, W., Mihalas, D., 
Hummer, D.G., \& Mihalas, B.W. 1988, ApJ, 332, 261

\bibitem[D\"appen et al. 1991]{da91}
D\"appen W., Gough D. O., Kosovichev A. G., \& Thompson M. J. 1991, 
in Lecture Notes in Physics, Vol. 388, ed., Gough D. O., Toomre J.
(Heidelberg: Springer), 111

\bibitem[D\"appen et al. 1993]{da93}
D\"{a}ppen, W., Gough, D.O., Kosovichev, A.G., \& 
Rhodes, E.J., Jr. 1993, 
%[On the influence of the treatment of heavy elements in the equation of 
%state on the resulting values of the adiabatic exponent $\gamma_1$], 
in Proc. IAU Symposium No 137: Inside the Stars, ed. 
W. Weiss \& A. Baglin (PASP Conference Series Vol. 40), 304

\bibitem[Dziembowski, Pamyatnykh \& Sienkiewicz 1990]{dps90}
Dziembowski W. A., Pamyatnykh A. A., \& Sienkiewicz R. 1990, 
MNRAS, 244, 542

\bibitem[Dziembowski, Pamyatnykh \& Sienkiewicz 1992]{dps92}
Dziembowski W. A., Pamyatnykh A. A., \& Sienkiewicz R. 1992,
Acta Astron., 42, 5

\bibitem[Dziembowski et al.~1994]{dzi94}
Dziembowski W. A., Goode P. R., Pamyatnykh A. A., \& Sienkiewich R., 
1994, ApJ, 432, 417

\bibitem[Ebeling et al.]{ekk76}
Ebeling, W., Kraeft, W.D., \& Kremp, D. 1976,
Theory of Bound States and Ionization Equilibrium 
in Plasmas and Solids (DDR-Berlin: Akademieverlag)

\bibitem[Ebeling et al.]{eb91}
Ebeling, W., F\"orster, A., Fortov, V.E., 
Gryaznov, V.K., \& Polishchuk, A.Ya. 1991,
Thermodynamic Properties of Hot Dense Plasmas
(Stuttgart: Teubner)

\bibitem[Eggleton, Faulkner \& Flannery (1973)]
{eff73}Eggleton, P. P., Faulkner, J., \& Flannery, B. P. 1973,
Astron. Astrophys., 23, 325

\bibitem[Elliot 1996]{el96}Elliot J.R. 1996, MNRAS, 280, 1244

\bibitem[Elliot \& Kosovichev]{ek98}Elliot J.R., \& Kosovichev, A.G 
1998, ApJ, 500, L199

\bibitem[Gabriel 1994]{g94}
Gabriel, M. 1994, Astron. Astrophys., 292, 281

\bibitem[Gough et al. 1996]{go96}Gough D. O., Kosovichev A. G., Toomre
J., \& the GONG Team 1996,
Science, 272, 1296

\bibitem[Grevesse \& Noels]{gn93}Grevesse N., Noels A. 1993,
in Origin and evolution of the Elements,
ed. Prantzos N., Vangioni-Flam E., \& Cass\'e M.,
(Cambridge: Cambridge University Press), 15

\bibitem[Guenther 1994]{gue94}
Guenther, D.B. 1994, ApJ, 422, 400

\bibitem[Harris et al. 1960]{ha60} 
Harris, G.M., Roberts, J.E., \& Trulio, J.G. 1960,
Phys. Rev., 119, 1832

\bibitem[Hummer \& Mihalas]{hm88} Hummer, D.G., \& Mihalas, D. 1988,
ApJ, 331, 794

\bibitem[Iglesias \& Rogers 1995]{ir95}
Iglesias, C.A., \& Rogers, F.J. 1995, ApJ, 443, 460

\bibitem[Iglesias \& Rogers 1996]{ir96}Iglesias, C. A., \& 
Rogers, F. J. 1996, ApJ, 464, 943

\bibitem[Kosovichev 1995]{ko95}
Kosovichev A. G. 1995 in Proc: Helio- and Asteroseismology from the
Earth and Space, ed. Ulrich, R. K., Rhodes, E. J., D\"appen, W.
(ASPCS Vol. 76), 89

\bibitem[Kosovichev 1996]{ko96}Kosovichev A. G. 1996, 
Bull. Astron. Soc. India, 24, 355

\bibitem[Kosovichev]{kos97}Kosovichev A. G. 1997, 
in Robotic Exploration close to the Sun: Scientific Basis,
ed. S.R. Habbal, AIP Conf. Proc. 385 (Woodbury, NY: Amer. Inst. Phys.), 159.

\bibitem[Kosovichev et al.~1997]{ko97}Kosovichev A. G., Schou J.,
Scherrer P., et al. 1997, Solar Phys., 170, 43

\bibitem[Kraeft et al.]{k86}
Kraeft W.D., Kremp, D., Ebeling, W., \& R\"opke G. 1986,
Quantum Statistics of Charged Particle Systems
(New York: Plenum)

\bibitem[Kurucz]{ku91}Kurucz R. L. 1991, 
in NATO ASI Series, Stellar Atmospheres: Beyond Classical Models,
ed. Crivellari L., Hubeny I., Hummer D.G. (Dordrecht: Kluwer), 441

\bibitem[Lubow et al. 1980]{l80}
Lubow, S.H., Rhodes, E.J., \& Ulrich, R.K. 1980,
%[Five minute oscillations as a probe of the solar interior].
in Lecture Notes in Physics,
Vol. 125: Nonradial and Nonlinear Stellar Pulsation,
ed. Hill, H.A. and Dziembowski, W. (Berlin: Springer), 300

\bibitem[Mihalas et al. 1988]
{mdh88} Mihalas, D., D\"appen, W., \& Hummer, D.G. 1988, 
ApJ, 332, 815

\bibitem[Nayfonov \& D\"appen]{nd98} 
Nayfonov, A., \& D\"appen, W. 1998,
ApJ, 499, 489

\bibitem[Noels et al. 1984]{nsg84}
Noels, A., Scuflaire, R., \& Gabriel, M. 1984,
%[Influence of the equation of state on the solar five-minute oscillation].
Astron. Astrophys.,
130, 389

\bibitem[Perez \& D\"appen 1998]{pd98} 
Perez, A., \& D\"appen, W. 1998, {\apj}, in preparation

\bibitem[Rhodes et al. 1997]{rho97}
Rhodes E. J., Kosovichev A. G., Schou J., 
Scherrer P. H., \& Reiter, J. 1997, Solar Phys., 175, 287

\bibitem[Richard et al.]{ric98}
Richard, O., Dziembowski, W.A., Sienkiewicz, R. \& Goode, P.R. 1998,
Astron. Astrophys., 338, 756.

\bibitem[Rogers 1977]{ro77}
Rogers, F.J. 1977,
Phys. Lett., 61A, 358

\bibitem[Rogers 1981]{ro81}
Rogers, F.J. 1981,
Phys. Rev., A24, 1531

\bibitem[Rogers 1986]{ro86}
Rogers, F.J. 1986, {\apj}, 310, 723

\bibitem[Rogers et al.]{rsi96}
Rogers, F.J., Swenson, F.J., \& Iglesias, C.A. 1996, {\apj},
456, 902

\bibitem[Rosenthal et al. 1995]{ros95}
Rosenthal, C.S., Christensen--Dalsgaard, J., Houdek, G.,
Monteiro, M., Nordlund, \AA., Trampedach, R, 1995
in Proc. of 4th SOHO Workshop, ESA-SP 376,
ed. J.T.Hoeksema, V. Domingo, B. Fleck,
B. Battrick (Noordwijk: ESA), 459 

\bibitem[Rosenthal et al. 1998]{ros98}
Rosenthal, C.S., Christensen--Dalsgaard, J.,
Nordlund, \AA., Stein, R., Trampedach, R, 1998
A\&A, submitted (astro-ph/9803206)

\bibitem[Schou et al. 1997]{sch97}Schou J., Kosovichev A. G., Goode P.
R., \& Dziembowski W. A. 1997, ApJ, 489, L197

\bibitem[Schou et al. 1998]{sch98}Schou J., Antia, H.M., 
Basu, S., et al. 1998, ApJ, 505, 390

\bibitem[Shibahashi et al. 1983; 1984]{sng83-84}
Shibahashi, H., Noels, A., \& Gabriel, M. 1983,
%[Influence of the equations of state and of the $Z$ value on the solar
%five-minute oscillation].
Astron. Astrophys.,
123, 283

%--- the following ref is really used and cited with the previous one
\bibitem[]{}
Shibahashi, H., Noels, A., \& Gabriel, M. 1984,
%[Influence of the equations of state and of the $Z$ value on the solar
%oscillations].
Mem. Soc. Astron. Ital., 55, 163

\bibitem[Thompson et~al.~1996]{tho96}Thompson M. J., Toomre J., 
Anderson E. R., \& the GONG Team 1996, Science, 272, 1300

\bibitem[Ulrich 1982]{ul82}
Ulrich, R.K. 1982,
%[The influence of partial ionization and scattering
%states on the solar interior structure].
ApJ,
258, 404

\bibitem[Ulrich \& Rhodes 1983]{ur83}
Ulrich, R.K., \& Rhodes, E.J. 1983,
%[Testing solar models with global solar oscillations in the 5-minute band].
ApJ,
265, 551

\end{thebibliography}
\end{document}